\documentclass[prd,preprint,onecolumn,showpacs,preprintnumbers,amsmath,nofootinbib,amssymb,floatfix]{revtex4-1}
\usepackage{graphicx}
\usepackage{amsmath}
\usepackage{amsfonts}
\usepackage{amssymb}
\usepackage{multirow}
\usepackage{color}
\usepackage{soul}
\usepackage{slashed}

\begin{document}
\title{Magnetic moments of the low-lying $J^P=\,1/2^-$, $3/2^-$  $\Lambda$ resonances within the framework of the chiral quark model}
\author{A. Mart\'inez Torres$^1$, K. P. Khemchandani$^1$, Neetika Sharma$^2$, Harleen Dahiya$^2$\\}
\affiliation{
$^1$Instituto de F\'isica, Universidade de S\~ao Paulo, C.P 66318, 05314-970 S\~ao Paulo, SP, Brazil.\\
$^2$Department of Physics, Dr. B. R. Ambedkar National \\ Institute of Technology, Jalandhar-144011, India.\\
}

\date{\today}

\begin{abstract}
The magnetic moments of the low-lying spin-parity $J^P=$ $1/2^-$, $3/2^-$ $\Lambda$ resonances, like, for example, $\Lambda(1405)$ $1/2^-$, $\Lambda(1520)$ $3/2^-$,
as well as their transition magnetic moments, are calculated using the chiral quark model. The results found are compared with those obtained from the nonrelativistic quark model and those of unitary chiral theories, where some of these states are generated through the dynamics of two hadron coupled channels and their unitarization.
\end{abstract}

\maketitle

\section{Introduction}
One of the most intriguing issues of the present day nuclear-particle physics is to elucidate the nature of hadron resonances using QCD based theories.  There are different ways to look into the underlying structure of a hadron and a lot of effort has been spent in this direction, specially, by studying radiative decays and electromagnetic properties of the ground state baryons and mesons and their resonances~\cite{Cheng,Linde,JidoMM,Chiang,Harleen,Liu,Yu,MiSo,JidoFF,Hideko,Daniel,Neetika,Geng1,Geng2}. These studies reveal that instead of the traditional picture of a baryon as a three quark state and a meson as a quark antiquark pair~\cite{IK}, some excited hadrons seem to have more complicated structures, where the contribution from the meson cloud or a picture in which the hadrons are the building blocks of the theory seem to take an advantage. Some examples of these type of hadrons are $\Lambda(1405)$,
$\Lambda(1670)$, $\Lambda(1520)$, $f_0(980)$, $a_0(980)$, $\phi (2170)$, $N^*(1710)$, etc. These states have been studied within effective field theories based on chiral Lagrangians and unitarity and it has been found that the observed properties, like mass, width, partial decay widths, etc., are well understood within a framework in which these states arise as a consequence of the underlying hadron dynamics when different relevant coupled channels are considered. For example, $\Lambda(1405)$ is generated in the $\bar KN$-$\pi\Sigma$ system, $\sigma(600)$, $f_0(980)$ and $a_o(980)$ in the $K\bar K$, $\pi\pi$, and $\pi\eta$ dynamics, $\phi(2170)$ is found to get formed in the $\phi K\bar K$ system, etc.,~\cite{JidoMM, MiSo,Kaiser, OllerM, JidoL,MKO1,MKO2, Luis}. In these approaches, the scattering matrix is obtained by solving the Bethe-Salpeter equation and poles found in the second Riemann sheet are related to resonant states. From the pole position the mass and width of the resonance is extracted directly and the coupling of the resonance to the different channels can be obtained from the residues of the scattering amplitude at the pole position. However, the evaluation of other static properties of the resonance, for example, magnetic moments, is not so straightforward, since the wave functions and operators are not manifestly present in these approaches. Thus, alternative methods involving the calculation of the scattering matrix are required~\cite{JidoMM,Hyodo}. This situation is in contrast with the one of the SU(6) quark models,  where the wave function of a resonance is given as a superposition of different spin states in the 70 dimensional representation and the magnetic moment is evaluated from the matrix elements of the magnetic moment operator for the given wave function.

A model which has the advantages of the conventional quark model in the evaluation of the electromagnetic properties of a resonance, but which goes beyond it, is the chiral quark model ($\chi$QM)~\cite{Weinberg,Manohar}. In this case, the dominant process is the fluctuation of a valence quark $q$ into a quark $q^\prime$ through the emission of a Goldstone boson. Hence, the
influence that the Goldstone boson degrees of freedom could have on the magnetic moment of a state is taken into account~\cite{Cheng, Linde,Harleen,Liu, Neetika}. However,
unlike the effective field theories based on chiral Lagrangians, no unitarization procedure is considered in the chiral quark model. It would be interesting to know what can be the results obtained within a chiral quark model for the magnetic moment of resonances like the $\Lambda(1405)$, $\Lambda(1670)$, etc., where the presence  of the meson cloud plays a crucial role in the determination of their properties. This is precisely the objective of this paper.

The paper is organized as follows: first, we introduce the nonrelativistic SU(6) quark model and within the same the method to calculate the magnetic moment of the low-lying $J^P=$ $1/2^-$, $3/2^-$ $\Lambda$ resonances and their transitions. Next,  we introduce the chiral quark model and the procedure needed to determine the effect of the presence of the Goldstone bosons on the magnetic moment of the $\Lambda$ resonances. Finally, we show the results found within the two models, and, whenever possible, we compare the results with the findings of unitary chiral theories (which are built on hadronic degrees of freedom) and draw some conclusions.

\section{Magnetic moments in the SU(6) quark model}\label{su6}
In the nonrelativistic SU(6) constituent quark model, the low-lying negative-parity $\Lambda$ resonances are described as $p$-wave excitations (thus total orbital angular momentum $L=1$) of the 70-dimensional representation, which has the following SU(2)$\times$SU(3) decomposition
\begin{align}
70={}^2 8\oplus{}^4 8\oplus {}^2 1\oplus {}^2 10.\label{70}
\end{align}
In Eq.~(\ref{70}) we have adopted a notation which mimics the usual spectroscopic notation, $^{2S+1} D$, to indicate the total quark spin $S=1/2, 3/2$ and the dimension $D$ of the flavor SU(3) representation.

The $\Lambda$ particles are isospin singlets, thus, their wave functions must be a linear combination of the flavor octet and singlet states, i.e.,
\begin{align}
|\Lambda\rangle=a_1|{}^2 8\rangle +a_2|{}^4 8\rangle+a_3|{}^2 1\rangle,\label{Lamb}
\end{align}
where the $a_1$, $a_2$, $a_3$ coefficients (different for each $\Lambda$ resonance) are defined such that the $|\Lambda\rangle$ state is normalized to unity, i.e., $a_1^2+a_2^2+a_3^2=1$.

The spin $S=1/2,3/2$ states present in Eq.~(\ref{70}) are coupled with the orbital angular momentum $L=1$ such that for a certain total angular momentum $J=L\oplus S$, the wave function of the $|{}^2 8\rangle$, $|{}^4 8\rangle$ and $|{}^2 1\rangle$ states in Eq.~(\ref{Lamb}) are given by~\cite{JidoMM,Chiang,Liu}
\begin{align}
|{}^2 8; JM\rangle&=\sum_{m_L}\sum_{m_S}C\left(1\, \frac{1}{2} \,m_L\, m_S\Big |J\,M\right)\frac{1}{2}\left\{\psi^\rho_{L m_L}\chi^\rho_{m_S}\phi^\lambda+\psi^\rho_{L m_L}\chi^\lambda_{m_S}\phi^\rho\right. \nonumber\\
&\quad\quad+\left. \psi^\lambda_{L m_L}\chi^\rho_{m_S}\phi^\rho-\psi^\lambda_{L m_L}\chi^\lambda_{m_S}\phi^\lambda\right\},
\\
|{}^4 8; JM\rangle&=\sum_{m_L}\sum_{m_S}C\left(1\, \frac{3}{2} \,m_L\, m_S\Big |J\,M\right)\frac{1}{\sqrt{2}}\left\{\psi^\lambda_{L m_L}\chi^S_{m_S}\phi^\lambda+\psi^\rho_{L m_L}\chi^S_{m_S}\phi^\rho\right\},
\\
|{}^2 1; JM\rangle&=\sum_{m_L}\sum_{m_S}C\left(1\, \frac{1}{2} \,m_L\, m_S\Big |J\,M\right)\frac{1}{\sqrt{2}}\left\{\psi^\lambda_{L m_L}\chi^\rho_{m_S}\phi_A-\psi^\rho_{L m_L}\chi^\lambda_{m_S}\phi_A\right\}.
\end{align}
Here $m_L$, $m_S$ and $M$ correspond to the projection of the orbital, spin and total angular momenta on the $z$-axis, respectively, $C(L\, S \,m_L \,m_S|J\, M) $ are Clebsch-Gordan coefficients, and $\psi$, $\chi$, $\phi$ represent  the spatial, spin, and flavor wave functions. The superscript $S$ ($A$) or $\rho$ ($\lambda$) in these wave functions indicate the symmetry related to the three quarks: totally symmetric (antisymmetric) among the three quarks or odd (even) under the exchange of the first two quarks. Following Ref.~\cite{IK}, the spin 1/2 and 3/2 wave functions are given as:
\begin{align}
&\chi^S_{\frac{3}{2}}=|\uparrow\,\uparrow\,\uparrow\rangle, \quad
\chi^S_{\frac{1}{2}}=\frac{1}{\sqrt{3}}\Big[|\uparrow\,\uparrow\,\downarrow\rangle+|\uparrow\,\downarrow\,\uparrow\rangle+|\downarrow\,\uparrow\,\uparrow\rangle\Big],\nonumber\\
&\chi^S_{-\frac{3}{2}}=|\downarrow\,\downarrow\,\downarrow\rangle,\quad
\chi^S_{-\frac{1}{2}}=\frac{1}{\sqrt{3}}\Big[|\downarrow\,\downarrow\,\uparrow\rangle+|\downarrow\,\uparrow\,\downarrow\rangle+|\uparrow\,\downarrow\,\downarrow\rangle\Big],\nonumber\\
&\chi^\rho_{\frac{1}{2}}=-\frac{1}{\sqrt{2}}\Big[|\uparrow\,\downarrow\,\uparrow\rangle-|\downarrow\,\uparrow\,\uparrow\rangle\Big],\quad\chi^\rho_{-\frac{1}{2}}=\frac{1}{\sqrt{2}}\Big[|\downarrow\,\uparrow\,\downarrow\rangle-|\uparrow\,\downarrow\,\downarrow\rangle\Big],\\
&\chi^\lambda_{\frac{1}{2}}=-\frac{1}{\sqrt{6}}\Big[2|\uparrow\,\uparrow\,\downarrow\rangle-|\uparrow\,\downarrow\,\uparrow\rangle-|\downarrow\,\uparrow\,\uparrow\rangle\Big],\quad\chi^\lambda_{-\frac{1}{2}}=\frac{1}{\sqrt{6}}\Big[2|\downarrow\,\downarrow\,\uparrow\rangle-|\downarrow\,\uparrow\,\downarrow\rangle-|\uparrow\,\downarrow\,\downarrow\rangle\Big]\nonumber,
\end{align}
and, for the $\Lambda$ states, the flavor wave functions  are:
\begin{align}
\phi^\rho&=\frac{1}{\sqrt{12}}[2uds-2dus+usd-dsu-sud+sdu],\nonumber\\
\phi^\lambda&=\frac{1}{2}[usd-dsu+sud-sdu],\\
\phi_A&=\frac{1}{\sqrt{6}}[uds-dus-usd+dsu+sud-sdu].\nonumber
\end{align}

In nonrelativistic quark models, as the one of Ref.~\cite{IK},
the orbital motion of the system is described by the three-dimensional harmonic oscillator wave functions, $\psi_{NLm_L}$~\cite{IK,Capstick}. In such a case, it is convenient to work with Jacobi coordinates, defined in terms of the three quark positions $\vec{r}_i$ as
\begin{align}
\vec{\rho}&=\frac{1}{\sqrt{2}}(\vec{r}_1-\vec{r}_2),\nonumber\\
\vec{\lambda}&=\frac{1}{\sqrt{6}}(\vec{r}_1+\vec{r}_2-2\vec{r}_3),\label{Jacobi}
\end{align}
since in terms of these variables the Hamiltonian of the three quark system gets separated into two independent three-dimensional oscillators. In this way, the orbital angular momentum is given by $\vec{L}=\vec{l}_\rho+\vec{l}_\lambda$, with $\vec{l}_\rho=\vec{\rho}\times \vec{p}_\rho$, $\vec{l}_\lambda=\vec{\lambda}\times \vec{p}_\lambda$, where $\vec{p}_\rho=-i\hbar\vec{\nabla}_\rho$ and $\vec{p}_\lambda=-i\hbar\vec{\nabla}_\lambda$. The ground state of a baryon consists of three quarks in the ground states, thus, $N=0$, while the first excited state, $N=1$, is realized when one of the three quarks is excited to the $P$-state with $L=1$, having either $l_\rho=1$ or $l_\lambda=1$. The orbital angular momenta $\vec{l}_\rho$ and $\vec{l}_\lambda$ operate over these spatial wave function as follows~\cite{Chiang}  (in the following we omit the index $N$ in the wave functions for simplicity)
\begin{align}
l^z_\rho\psi^\rho_{1 m_L}&=m_L\psi^\rho_{1 m_L},\quad l^z_\rho\psi^\lambda_{1 m_L}=0,\nonumber\\
l^z_\lambda\psi^\lambda_{1 m_L}&=m_L\psi^\lambda_{1 m_L},\quad l^z_\lambda\psi^\rho_{1 m_L}=0,\nonumber\\
(\vec{\rho}\times\vec{p}_\lambda)^z\psi^\lambda_{1 m_L}&=m_L\psi^\rho_{1 m_L},\quad(\vec{\rho}\times\vec{p}_\lambda)^z\psi^\rho_{1 m_L}=0,\\
(\vec{\lambda}\times\vec{p}_\rho)^z\psi^\rho_{1 m_L}&=m_L\psi^\lambda_{1 m_L},\quad(\vec{\lambda}\times\vec{p}_\rho)^z\psi^\lambda_{1 m_L}=0,\nonumber
\end{align}
where the superscript $z$ indicates the $z$ component of the vector.

In the nonrelativistic quark model, the magnetic moment operator for a baryon $B$ is the sum of the contributions coming from the spin and orbital angular momenta of the three quarks which constitute the baryon,
\begin{align}
\vec{\mu}_B=\vec{\mu}^S_B+\vec{\mu}^L_B,\quad\vec{\mu}^S_B=\sum_{i=1}^{3}\mu_i\vec{\sigma}(i),\quad \vec{\mu}^L_B=\sum_{i=1}^{3}\mu_i \vec{L}(i),\label{mag2}
\end{align}
with $\mu_i=Q_i/(2M_i)$ being the quark magnetic moment, where $Q_i$ and $M_i$ are the charge and mass of the {\it ith} quark, respectively, $\vec{\sigma}$ the Pauli matrices and $\vec{L}(i)=\vec{r}_i\times\vec{p}_i$ can be written in terms of $\vec{l}_\rho$, $\vec{l}_\lambda$, $\vec{\rho}\times\vec{p}_\lambda$ and $\vec{\lambda}\times\vec{p}_\rho$ using Eq.~(\ref{Jacobi}) ~\cite{Chiang}.
To determine the magnetic moment of the low-lying $1/2^-$ and $3/2^-$ $\Lambda$ resonances, we need to calculate the expectation value of the {\it zth} component of the operator $\vec{\mu}_B$, $\mu_{Bz}$, between the state of Eq.~(\ref{Lamb}) for a certain total angular momentum $J$ and its third component $M$, that is,
\begin{equation}
\mu_\Lambda(J,M)\equiv\langle\Lambda; J,M|\mu_{Bz}|\Lambda;J,M\rangle,
\end{equation}
where, from Eq.~(\ref{mag2})
\begin{align}
\mu_\Lambda(J,M)&=\mu^S_{\Lambda\,z}(J,M)+\mu^L_{\Lambda\,z}(J,M),\nonumber\\
\mu^S_{\Lambda\,z}(J,M)&\equiv\langle\Lambda; J,M|\mu^S_z|\Lambda;J,M\rangle,\quad\mu^S_z=\sum_{i=1}^{3}\mu_i\sigma_z(i), \nonumber\\
\mu^L_{\Lambda\,z}(J,M)&\equiv\langle\Lambda; J,M|\mu^L_z|\Lambda;J,M\rangle,\quad\mu^L_z \sum_{i=1}^{3}\mu_i L_z(i).\label{muL}
\end{align}
For a certain total angular momentum $J$, defining the state $|{}^28\rangle$ as $|1\rangle$, $|{}^4 8\rangle$ as $|2\rangle$ and $|{}^2 1\rangle$ as $|3\rangle$, and using Eq.~(\ref{Lamb}), we get
\begin{align}
\mu^S_{\Lambda\,z}(J,M)&=\sum_{m=1}^3\sum_{n=1}^3 a_m a_n \langle JM|\mu^S_z|JM\rangle_{m\,n}\label{muS},\\
\mu^L_{\Lambda\,z}(J,M)&=\sum_{m=1}^3\sum_{n=1}^3 a_m a_n \langle JM|\mu^L_z|JM\rangle_{m\,n}\label{muL},
\end{align}
Thus, the calculation of the magnetic moment for the low-lying $\Lambda$ resonances reduces to the determination of the transition matrix elements $\langle JM|\mu^S_z| JM\rangle$ and $\langle JM|\mu^L_z|JM\rangle$. In Eqs.~(\ref{S1h1h})-(\ref{L3h}) we show the results found for these transitions matrix elements for $(J,M)=(1/2,1/2), (3/2,1/2), (3/2,3/2)$, where we have defined
$A\equiv\mu_u+\mu_d+\mu_s$ and $B=\mu_u+\mu_d-2\mu_s$\footnote{The results found here for $J=1/2, M=1/2$ are same as the ones obtained by the authors of Ref.~\cite{JidoMM}. Note, however, that within their normalization for the spin and flavor wave functions, there should be a global minus sign in the transition matrix elements $\langle{}^4 8|\mu^S_z|{}^2 1\rangle$, $\langle{}^2 1|\mu^S_z|{}^4 8\rangle$ and the coefficient $a_2$ should be replaced by $-a_2$ in order to be consistent. A factor $2/3$ should be multiplied to the orbital magnetic moment transition elements. The sign as well as the $2/3$ factor are missing in their calculations. We thank A. Hosaka for clarifying this issue.}. The element $(1,1)$ in the matrices corresponds to the transition between the states ${}^28\to {}^28$, the element $(1,2)$ to the transition ${}^2 8\to {}^48$, etc. As it can be seen, for the orbital angular momentum contribution, the matrix element involving the transition ${}^2 8\to {}^48$, and vice versa, is zero. This is because the spin parts of the wave functions of the states $|{}^2 8;JM\rangle$ and $|{}^48;JM\rangle$ are orthogonal, which remain unaltered by the orbital angular momentum operator.
\begin{align}
\left\langle\frac{1}{2}\frac{1}{2} \Big|\mu^S_z\Big|\frac{1}{2}\frac{1}{2}\right \rangle&=\frac{1}{9}\left (\begin{array}{ccc} -A&-2B&B\\-2B&5A&2B\\B&2B&-A\end{array}\right),\label{S1h1h}\\
\nonumber\\
\left\langle\frac{3}{2}\frac{1}{2} \Big|\mu^S_z\Big|\frac{3}{2}\frac{1}{2}\right \rangle&=\frac{1}{3}\left\langle\frac{3}{2}\frac{3}{2} \Big|\mu^S_z\Big|\frac{3}{2}\frac{3}{2}\right \rangle=\dfrac{1}{9}\left (\begin{array}{ccc} A&-\sqrt{\dfrac{2}{5}}B&-B\\-\sqrt{\dfrac{2}{5}}B&\dfrac{11}{5}A&\sqrt{\dfrac{2}{5}}B\\-B&\sqrt{\dfrac{2}{5}}B&A\end{array}\right),\label{S3h}
\end{align}

\begin{align}
\left\langle\frac{1}{2}\frac{1}{2} \Big|\mu^L_z\Big|\frac{1}{2}\frac{1}{2}\right \rangle&=\frac{1}{9}\left (\begin{array}{ccc} 2A&0&-B\\0&-\dfrac{2A+B}{2}&0\\-B&0&2A\end{array}\right),\label{L1h1h}\\
\nonumber\\
\left\langle \frac{3}{2}\frac{1}{2} \Big|\mu^L_z\Big|\frac{3}{2}\frac{1}{2}\right \rangle&=\frac{1}{3}\left\langle\frac{3}{2}\frac{3}{2} \Big|\mu^L_z\Big|\frac{3}{2}\frac{3}{2}\right \rangle=\frac{1}{9}\left (\begin{array}{ccc} A&0&-\dfrac{B}{2}\\0&\dfrac{2A+B}{5}&0\\-\dfrac{B}{2}&0&A\end{array}\right).\label{L3h}
\end{align}

Each of the elements in the matrices of Eqs.~(\ref{S1h1h})-(\ref{L3h}) can be written as
\begin{align}
\sum_{q=u,d,s}\Delta q^S_{\textrm{val}}\mu_q,
\end{align}
for the spin contributions, and
\begin{align}
\sum_{q=u,d,s}\Delta q^L_{\textrm{val}}\mu_q,
\end{align}
for the orbital contributions, with $\Delta q^S_{\textrm{val}}=q^+-q^-$ and $\Delta q^L_{\textrm{val}}=q^{(+1)}-q^{(-1)} $ the spin and orbital polarizations, respectively, of the valence quarks for the transition matrix element considered. Here, $q^+$ $(q^-)$ represents the number of quarks with spin up (down) and $q^{(+1)}$ ($q^{(-1)})$ stands for the number of quarks with the projection of the orbital angular momentum being $+1$ ($-1$).

Note that Eqs.~(\ref{muS}) and (\ref{muL}) correspond to the magnetic moment of a certain $\Lambda$ resonance. However, it is also possible to determine the magnetic moment
for a transition involving two different $\Lambda$ states. The wave functions related with these two $\Lambda$ resonances will have the form of Eq.~(\ref{Lamb}), but
different coefficients in front of the states $|{}^2 8\rangle$, $|{}^4 8\rangle$,$|{}^2 1\rangle$. Thus, if we call these resonances as $\Lambda$ and $\Lambda^\prime$, respectively, the wave functions for these two states will be of the form
\begin{align}
|\Lambda\rangle&=a_1|{}^2 8\rangle+a_2|{}^4 8\rangle+a_3|{}^2 1\rangle,\\
|\Lambda^\prime\rangle&=a^\prime_1 |{}^2 8\rangle+a^\prime_2 |{}^4 8\rangle+a^\prime_3 |{}^2 1\rangle,
\end{align}
and thus, analogously to Eqs.~(\ref{muS}) and (\ref{muL}), the spin and orbital transition magnetic moments between these resonances will be given by
\begin{align}
\mu^S_{\Lambda\to\Lambda^\prime\,z}(J,M)&=\sum_{m=1}^3\sum_{n=1}^3 a_m a^\prime_n \langle m;JM|\mu^S_z|n;JM\rangle,\label{muineS}\\
\mu^L_{\Lambda\to\Lambda^\prime\,z}(J,M)&=\sum_{m=1}^3\sum_{n=1}^3 a_m a^\prime_n \langle m;JM|\mu^L_z|n;JM\rangle.\label{muineL}
\end{align}
\section{Magnetic moments in the chiral quark model}
The basic idea of the chiral quark model~\cite{Weinberg,Manohar} is that the nonperturbative QCD phenomenon of chiral symmetry breaking ($\chi$SB) takes place at a distance scale significantly smaller than that of color confinement ( $\Lambda_{\chi \textrm{SB}}\sim1$ GeV while experimental data indicate that the confinement scale is $\Lambda_{\textrm{QCD}}=100-300$ MeV). Thus, in the interior of a hadron, in the scale range between $\Lambda_{\chi\textrm{QCD}}$ and $\Lambda_{\chi \textrm{SB}}$, the effective degrees of freedom are the constituent quarks and the Goldstone bosons (GBs). Thus, properties of a hadron, like spin, magnetic moment, etc., can be understood by the presence of a quark sea generated by the emission of GBs from the constituent quarks of the hadron (valence quarks). This emission of GBs creates quark-antiquark pairs with quantum numbers $J^P=0^-$ from the vacuum and flips the quark spin direction~\cite{Cheng,Linde,Harleen, Liu, Neetika,Sharma,Song},
\begin{align}
q^{\pm}\to {q^\prime}^\mp+\textrm{GB}\to{q^\prime}^\mp+(q\bar q^\prime).\label{tran}
\end{align}

In the $\chi$QM, the effective Lagrangian describing the interaction between the quarks and the Goldstone bosons is given by
\begin{align}
\mathcal{L}=g_8\bar q \phi q,
\end{align}
where $q=\left(\begin{array}{c}u\\d\\s\end{array}\right)$ and $\phi$ is a matrix containing the Goldstone bosons,
\begin{align}
\phi&=\left(\begin{array}{ccc}\dfrac{\pi^0}{\sqrt{2}}+\beta\dfrac{\eta}{\sqrt{6}}+\zeta\dfrac{\eta^\prime}{\sqrt{3}}&\pi^+&\alpha K^+\\
\pi^-&-\dfrac{\pi^0}{\sqrt{2}}+\beta\dfrac{\eta}{\sqrt{6}}+\zeta\dfrac{\eta^\prime}{\sqrt{3}}&\alpha K^0\\
\alpha K^-&\alpha \bar K^0&-\beta\dfrac{2\eta}{\sqrt{6}}+\zeta\dfrac{\eta^\prime}{\sqrt{3}}\end{array}\right).\label{phi}
\end{align}
In Eq.~(\ref{phi}), $\zeta\equiv g_1/g_8$, with $g_1$ and $g_8$ being the coupling constants for the singlet and octet of GBs, respectively. The parameters $\alpha$, $\beta$ and $\zeta$ are suppression factors to take into account SU(3) breaking effects due to the fact that the quark $s$ is heavier than the quarks $u$ and $d$. Since the quark $s$ (or/and antiquark $\bar s$) is present in the particles $K$, $\eta$ and $\eta^\prime$, the corresponding fields in Eq.~(\ref{phi}) are multiplied by the factors $\alpha$ (for $K$), $\beta$ (for $\eta$) and $\zeta$ (for $\eta^\prime$). The fields $\eta$ and $\eta^\prime$ are multiplied by different factors to consider the breaking of the U(3) symmetry. The parameter $a\equiv |g_8|^2$ denotes the probability of the chiral fluctuation
$u^{\pm} (d^{\,\pm})\to d^{\,\mp}(u^\mp)+\{\pi^+, \pi^-\}$, whereas $\alpha^2 a$, $\beta^2 a$ and $\zeta^2 a$ represent the probabilities of the fluctuations $u^\pm (d^\pm)\to s^{\mp}+\{K^-,K^0\}$,
$u^\pm (d^{\,\pm} , s^\pm)\to u^\mp (d^{\,\mp} , s^\mp)+\eta$, and $u^\pm (d^{\,\pm} , s^\pm)\to u^\mp (d^{\,\mp} , s^\mp)+\eta^\prime$, respectively.

Within this framework, the part of the magnetic moment of a certain baryon arising from the spin angular momentum will have, thus, contributions from the valence quarks, $\mu^S_{\textrm{val}}$, sea quarks, $\mu^S_{\textrm{sea}}$, as well as from the orbital angular momentum of the quark sea (${q^\prime}+(q\bar q^\prime)$ in Eq.~(\ref{tran})), $\mu^S_{\textrm{orbit}}$,
\begin{align}
\mu^S=\mu^S_{\textrm{val}}+\mu^S_{\textrm{sea}}+\mu^S_{\textrm{orbit}}.\label{muspin}
\end{align}
The quantity $\mu^S_{\textrm{val}}$ has its origin in the spin polarization of the constituent quarks and, therefore, corresponds to the results obtained for the spin part of the magnetic moment within the nonrelativistic quark model explained in the previous section.  The sea quark spin contribution, $\mu^S_{\textrm{sea}}$, can be calculated by substituting for each valence quark
\begin{align}
q^{\pm}\to-\sum_{\textrm{GB}} P_{[q,\textrm{GB}]} q^{\pm}+|\psi(q^\pm)|^2,
\end{align}
where $P_{[q,\textrm{GB}]}$ is the probability of emission of a Goldstone boson from the quark $q$ and $|\psi(q^\pm)|^2$ the probability of transforming a $q^\pm$ quark as in Eq.~(\ref{tran}). These quantities are given by~\cite{Linde,Harleen, Liu,Neetika}
\begin{align}
\sum_{\textrm{GB}}P_{[u,\textrm{GB}]} &=\frac{a}{6}\left(9+6\alpha^2+\beta^2+2\zeta^2\right),\nonumber\\
\sum_{\textrm{GB}}P_{[d,\textrm{GB}]} &=\frac{a}{6}\left(9+6\alpha^2+\beta^2+2\zeta^2\right),\label{psiGB}\\
\sum_{\textrm{GB}}P_{[s,\textrm{GB}]} &=\frac{a}{3}\left(6\alpha^2+2\beta^2+\zeta^2\right),\nonumber
\end{align}
and
\begin{align}
|\psi(u^\pm)|^2&=a\left[\frac{1}{6}\left(3+\beta^2+2\zeta^2\right)u^\mp+d^\mp+\alpha^2 s^\mp\right],\nonumber\\
|\psi(d^\pm)|^2&=a\left[u^\mp+\frac{1}{6}\left(3+\beta^2+2\zeta^2\right)d^\mp+\alpha^2 s^\mp\right],\label{psisq}\\
|\psi(s^\pm)|^2&=a\left[\alpha^2u^\mp+\alpha^2d^\mp+\frac{1}{3}\left(2\beta^2+\zeta^2\right) s^\mp\right].\nonumber
\end{align}
Equations ~(\ref{psiGB}) and (\ref{psisq}) clearly show that the process of Eq.~(\ref{tran}) changes, for the wave function considered, the spin structure with respect to the one associated with the valence quarks. This difference is defined as $ \Delta q^S_{\textrm{sea}}$, such that the contribution to the magnetic moment from the sea quarks is written as
\begin{align}
\mu^S_{\textrm{sea}}=\sum_{q=u,d,s} \Delta q^S_{\textrm{sea}}\,\mu_q.
\end{align}

As discussed in Ref.~\cite{Cheng}, the quark sea generated in Eq.~(\ref{tran}) carries a significant amount of orbital angular momentum. In fact, parity and angular momentum conservation imply that the final state quark $q^\prime$ and $(\bar{q}^\prime q)$ in the GB emission process of Eq.~(\ref{tran}) must be in a relative P-wave state, generating in this way a contribution to the magnetic moment, $\mu^S_{\textrm{orbit}}$. The orbital moment of each process $q^\pm\to {q^\prime}^\mp+\textrm{GB}$ is~\cite{Cheng,Harleen, Neetika,Sharma}
\begin{align}
{\mu(q^\pm\to {q^\prime}^\mp)}_L=\frac{Q_{q^\prime}}{2M_q}\langle l_{q\,z}\rangle+\frac{Q_q-Q_{q^\prime}}{2M_{\textrm{GB}}}\langle l_{\textrm{GB}\,z}\rangle,\label{muLto}
\end{align}
where the one unit of angular momentum is shared by the two bodies, i.e., $q^\prime$ and GB,
\begin{align}
\langle l_{q\,z}\rangle=\frac{M_{\textrm{GB}}}{M_q+M_{\textrm{GB}}},\quad\langle l_{\textrm{GB}\,z}\rangle=\frac{M_q}{M_q+M_{\textrm{GB}}},\label{lz}
\end{align}
The orbital moment in Eq.~(\ref{muLto}) is then multiplied by the probability for such a process to take place (which can be directly read from Eq.~(\ref{psisq})), to yield the magnetic moment due to all the transitions starting with a given valence quark:
\begin{align}
[\mu(u^\pm\to~)]&=\pm\, a\left[\frac{1}{6}\left(3+\beta^2+2\zeta^2\right)\mu(u^\pm\to u^\mp)+\mu(u^\pm\to d^\mp)+\alpha^2\mu(u^\pm\to s^\mp)\right],\nonumber\\
[\mu(d^\pm\to~)]&=\pm\, a\left[\mu(d^\pm\to u^\mp)+\frac{1}{6}\left(3+\beta^2+2\zeta^2\right)\mu(d^\pm\to d^\mp)+\alpha^2\mu(d^\pm\to s^\mp)\right],\label{muqwhatever}\\
[\mu(s^\pm\to~)]&=\pm\, a\left[\alpha^2\mu(s^\pm\to u^\mp)+\alpha^2\mu(s^\pm\to d^\mp)+\frac{1}{3}\left(2\beta^2+\zeta^2\right)\mu(s^\pm\to s^\mp)\right].\nonumber
\end{align}
Using Eqs.~(\ref{muLto}) and~(\ref{lz}), we can write Eq.~(\ref{muqwhatever}) in terms of the quark and Goldstone boson masses and the parameters of the $\chi$QM, i.e., $a$, $\alpha$, $\beta$ and $\zeta$, as
\begin{align}
[\mu(u^\pm\to~)]&=\pm\, a\left[\frac{3 M^2_u}{2 M_\pi (M_u+M_\pi)}-\frac{\alpha^2(M^2_K-3M^2_u)}{2M_K(M_u+M_K)}+\frac{\beta^2 M_\eta}{6(M_u+M_\eta)}+\frac{\zeta^2 M_{\eta^\prime}}{3(M_u+M_{\eta^\prime})}\right]\mu_u,\nonumber\\
[\mu(d^\pm\to~)]&=\mp\, 2a\left[\frac{3 (M^2_\pi-2M^2_d)}{4 M_\pi (M_d+M_\pi)}-\frac{\alpha^2 M_K}{2(M_d+M_K)}-\frac{\beta^2 M_\eta}{12(M_d+M_\eta)}-\frac{\zeta^2 M_{\eta^\prime}}{6(M_d+M_{\eta^\prime})}\right]\mu_d,\\
[\mu(s^\pm\to~)]&=\mp\, 2a\left[\frac{\alpha^2 (M^2_K-3M^2_s)}{2 M_K (M_s+M_K)}-\frac{\beta^2 M_\eta}{3(M_s+M_\eta)}-\frac{\zeta^2 M_{\eta^\prime}}{6(M_s+M_{\eta^\prime})}\right]\mu_s.\nonumber
\end{align}
Then, the contribution $\mu^S_{\textrm{orbit}}$ is given by
\begin{align}
\mu^S_{\textrm{orbit}}=\sum_{q=u,d,s}\Delta q^S_{\textrm{val}}\,\mu(q^+\to~).\label{muSto}
\end{align}

In the chiral quark model, the fraction of the magnetic moment of a baryon related with the orbital angular momentum, $\mu^L$, has contributions from the valence quarks, $\mu^L_{\textrm{val}}$, as well as from the sea quarks, $\mu^L_{\textrm{sea}}$. In other words:
\begin{align}
\mu^L=\mu^L_{\textrm{val}}+\mu^L_{\textrm{sea}}.\label{muangular}
\end{align}
The determination of $\mu^L_{\textrm{val}}$ corresponds to the calculation of the orbital contribution to the magnetic moment by the constituent quarks of the baryon in the nonrelativistic SU(6) model. This has already been discussed in Sec.~\ref{su6}, and the results are given by Eqs.~(\ref{L1h1h}) and (\ref{L3h}).

Similarly to the evaluation of the spin part of the magnetic moment, the sea quark orbital contribution can be obtained by using the following replacement in the wave function considered
\begin{align}
q^{(\pm 1)}\to -\sum_{\textrm{GB}}T_{[q,\textrm{GB}]} q^{(\pm 1)}+\left|\psi(q^{(\pm 1)})\right|^2\label{chiQL},
\end{align}
with $T_{[q,\textrm{GB}]}$ the probability of emitting a Goldstone boson from a quark $q^{(\pm 1)}$ and $\left|\psi(q^{(\pm 1)})\right|^2$ the probability of transforming a quark $q^{(\pm 1)}$ into a quark ${q^{\prime}}^{(\pm 1)}$~\cite{Liu}. These probabilities are given by,
\begin{align}
\sum_{\textrm{GB}}T_{[u,\textrm{GB}]}&=a(1+\alpha^2),\\
\sum_{\textrm{GB}}T_{[d,\textrm{GB}]}&=a(1+\alpha^2),\\
\sum_{\textrm{GB}}T_{[s,\textrm{GB}]}&=2a\alpha^2,
\end{align}
and 
\begin{align}
\left|\psi(u^{(\pm 1)})\right|^2&=a[d^{(\pm 1)}+\alpha^2 s^{(\pm 1)}],\nonumber\\
\left|\psi(d^{(\pm 1)})\right|^2&=a[u^{(\pm 1)}+\alpha^2 s^{(\pm 1)}],\\
\left|\psi(s^{(\pm 1)})\right|^2&=a\alpha^2[u^{(\pm 1)}+d^{(\pm 1)}].\nonumber
\end{align}
Equation~(\ref{chiQL}) alters the orbital quark polarizations associated with the valence quarks, giving rise to additional terms. This difference is defined as $\Delta q^L_{\textrm{sea}}$ and contributes to the orbital part of the magnetic moment of a certain baryon as
\begin{align}
\mu^L_{\textrm{sea}}=\sum_{q=u,d,s} \Delta q^L_{\textrm{sea}}\mu_q.
\end{align}

The quantities $\mu^S_{\textrm{sea}}$, $\mu^S_{\textrm{orbit}}$ need to be determined for the different matrix elements of Eqs.~(\ref{S1h1h}) and (\ref{S3h}), whereas $\mu^L_{\textrm{sea}}$
has to be calculated for the matrix elements of Eqs.~(\ref{L1h1h}) and (\ref{L3h}), i.e., for the different transitions involving the states $|{}^2 8\rangle$, $|{}^4 8\rangle$ and $|{}^2 1\rangle$ for a certain $J$ and $M$. In this way, we will obtain the matrix elements $\langle JM|\mu^S|JM\rangle$ and $\langle JM|\mu^L|JM\rangle$ associated with the chiral quark model and, using Eqs.~(\ref{muS}),~(\ref{muL}),~(\ref{muineS}), and~(\ref{muineL}), the magnetic moment for the different low-lying $\Lambda$ resonances and their transitions.

To calculate these matrix elements, first, we need to establish the value of the parameters present in the $\chi$QM, i.e., $a$, $\alpha$, $\beta$ and $\zeta$, which, as mentioned above, are related to the probability of fluctuation of a constituent quark by emitting Goldstone bosons.  These parameters are fixed in our case to the following values:
\begin{align}
a=0.12,\quad \alpha=\beta=0.45,\quad\zeta=-0.15,
\end{align}
which reproduce quite well the magnetic moments of the $1/2^+$ and $3/2^+$ baryons~\cite{Cheng,Harleen,Neetika,Sharma}. In the $\chi$QM, the quark and Goldstone boson masses also
enter in the evaluation of the magnetic moments. We use the following values for them:
\begin{align}
M_u=M_d=330,\quad M_s=510,\quad M_\pi=137, \quad M_K=496, \quad M_\eta=547, \quad M_{\eta^\prime}=958,
\end{align}
 all of them expressed in units of MeV. In this way, we have that
 \begin{align}
 \mu_u=\frac{Q_u}{2M_u}\cong2\mu_N,\quad\mu_d=\frac{Q_d}{2M_d}\cong -\mu_N,\quad \mu_s=\frac{Q_s}{2M_s}\cong -\frac{2}{3}\mu_N,
 \end{align}
where $\mu_N$ is the nuclear magneton.

Now we have all the ingredients necessary to calculate the transition matrix elements of Eqs.~(\ref{S1h1h})-(\ref{L3h}) using the chiral quark model. In the following we give the results obtained for the contributions $\mu^S_\textrm{sea}$ and $\mu^L_\textrm{sea}$. Introducing the quantities
\begin{align}
\cal{A}&=\frac{a}{27}\left[(\mu_u+\mu_d)(6\alpha^2+\beta^2+2\zeta^2+9)+2\mu_s(6\alpha^2+2\beta^2+\zeta^2)\right],\nonumber\\
\cal{B}&=-\frac{2a}{27}\left[(\mu_u+\mu_d)(3\alpha^2-\beta^2-2\zeta^2-9)+2\mu_s(3\alpha^2+4\beta^2+2\zeta^2)\right],\nonumber
\end{align}
we have,
\begin{align}
\left\langle\frac{1}{2}\frac{1}{2} \Big|\mu^S_\textrm{sea}\Big|\frac{1}{2}\frac{1}{2}\right \rangle&=\left(
\begin{array}{cccc}
  \cal{A} & \cal{B} & -\dfrac{\cal{B}}{2} \\
\cal{B} & -5\cal{A} &-\cal{B} \\
-\dfrac{\cal{B}}{2} & -\cal{B} & \cal{A} \\
\end{array}
\right),\label{S1h1hsea}\\\nonumber\\
\left\langle\frac{3}{2}\frac{1}{2} \Big|\mu^S_\textrm{sea}\Big|\frac{3}{2}\frac{1}{2}\right \rangle&=\frac{1}{3}\left\langle\frac{3}{2}\frac{3}{2} \Big|\mu^S_\textrm{sea}\Big|\frac{3}{2}\frac{3}{2}\right \rangle=\left(
\begin{array}{cccc}
   -\cal{A}& \dfrac{\cal{B}}{\sqrt{10}}& \dfrac{\cal{B}}{2} \\\\
\dfrac{\cal{B}}{\sqrt{10}} & -\dfrac{11}{5}\cal{A} & -\dfrac{\cal{B}}{\sqrt{10}} \\\\
 \dfrac{\cal{B}}{2} & -\dfrac{\cal{B}}{\sqrt{10}} &  -\cal{A}
\end{array}\label{S3hsea}
\right),
\end{align}
\begin{align}
\left\langle\frac{1}{2}\frac{1}{2} \Big|\mu^L_\textrm{sea}\Big|\frac{1}{2}\frac{1}{2}\right \rangle&=\frac{a\alpha^2}{3}\left(
\begin{array}{cccc}
0 & 0 & B \\
 0 & \dfrac{B}{2} & 0 \\
 B & 0 & 0 \\
\end{array}
\right),\label{L1h1hsea}\\
\left\langle\frac{3}{2}\frac{1}{2} \Big|\mu^L_\textrm{sea}\Big|\frac{3}{2}\frac{1}{2}\right \rangle&=\frac{1}{3}\left\langle\frac{3}{2}\frac{3}{2} \Big|\mu^L_\textrm{sea}\Big|\frac{3}{2}\frac{3}{2}\right \rangle=\frac{a\alpha^2}{6}\left(
\begin{array}{cccc}
0 & 0 & B \\
 0 & -\dfrac{2}{5}B & 0 \\
 B & 0 & 0 \\
\end{array}
\right).\label{L3hsea}
\end{align}
Using Eqs.~(\ref{S1h1h})-(\ref{L3h}) for the respective valence contributions, Eqs.~(\ref{S1h1hsea})-(\ref{L3hsea}) for the spin and orbital sea contributions, Eq.~(\ref{muSto}) for the Cheng-Li mechanism and Eqs.~(\ref{muspin}) and (\ref{muangular}), we find the following numerical results, in units of the nuclear
magneton, for $\mu^S$ and $\mu^L$ within the chiral quark model:
\begin{align}
\left\langle\frac{1}{2}\frac{1}{2} \Big|\mu^S\Big|\frac{1}{2}\frac{1}{2}\right \rangle&=\left(
\begin{array}{cccc}
  -0.003 & -0.445 & 0.222 \\
 -0.445 & 0.014 & 0.445 \\
 0.222 & 0.445 & -0.003 \\
\end{array}
\right),\\
\left\langle\frac{3}{2}\frac{1}{2} \Big|\mu^S\Big|\frac{3}{2}\frac{1}{2}\right \rangle&=\frac{1}{3}\left\langle\frac{3}{2}\frac{3}{2} \Big|\mu^S\Big|\frac{3}{2}\frac{3}{2}\right \rangle=\left(
\begin{array}{cccc}
   0.003 & -0.141 & -0.222 \\
 -0.141 & 0.006 & 0.141 \\
 -0.222 & 0.141 & 0.003 \\
\end{array}
\right),
\end{align}
\begin{align}
\left\langle\frac{1}{2}\frac{1}{2} \Big|\mu^L\Big|\frac{1}{2}\frac{1}{2}\right \rangle&=\left(
\begin{array}{cccc}
 0.074 & 0 & -0.240 \\
 0 & -0.157 & 0 \\
 -0.240 & 0 & 0.074 \\
\end{array}
\right),\\
\left\langle\frac{3}{2}\frac{1}{2} \Big|\mu^L\Big|\frac{3}{2}\frac{1}{2}\right \rangle&=\frac{1}{3}\left\langle\frac{3}{2}\frac{3}{2} \Big|\mu^L\Big|\frac{3}{2}\frac{3}{2}\right \rangle=\left(
\begin{array}{cccc}
0.037 & 0 & -0.120 \\
 0 & 0.063 & 0 \\
 -0.120 & 0 & 0.037 \\
\end{array}
\right).
\end{align}
\section{SU(6) quark model vs. Chiral quark model}
Once we have determined the transition matrix elements involving the states $|{}^2 8\rangle$, $|{}^4 8\rangle$ and $|{}^2 1\rangle$ for $J=1/2$, and $3/2$ within the SU(6) quark model as well as using the chiral quark model, we can proceed further and calculate the magnetic moments of the low-lying $\Lambda$ resonances and their transitions using Eqs.~(\ref{muS}),~(\ref{muL})~(\ref{muineS}),~(\ref{muineL}). For this, we need to know the coefficients $a_1$, $a_2$ and $a_3$ present in these equations and which give the weight of each of the  $|{}^2 8\rangle$, $|{}^4 8\rangle$ and $|{}^2 1\rangle$ states in the wave function of the $\Lambda$ resonances considered. Here we use for these coefficients the values obtained by Isgur and Karl~\cite{IK}, which are in good agreement with experimental findings in the strangeness 0 and -1 sectors.

\begin{table}[h!]
\caption{Magnetic moments in the SU(6) and chiral quark models for the low-lying $1/2^-$ and $3/2^-$ $\Lambda$ resonances.}\label{magmomres}
\begin{ruledtabular}
\begin{tabular}{c| cccc |ccc|cc}
&\multicolumn{4}{c|}{Spin}&\multicolumn{3}{c|}{Orbital}&Quark model&$\chi$QM\\
$\mu_{J,M}(\Lambda\to \Lambda^\prime)$&Valence&Sea&Orbit&Total&Valence&Sea&Total&Total&Total\\
\hline
 $\mu _{\frac{1}{2},\frac{1}{2}}(1405\to 1405) $& 0.194 & -0.006 & 0.006 & 0.194 &
   -0.128 & 0.015 & -0.113 & 0.066 & 0.081 \\
 $\mu _{\frac{1}{2},\frac{1}{2}}(1405\to 1670) $& 0.245 & -0.051 & 0.010 & 0.204 &
   -0.139 & 0.010 & -0.130 & 0.105 & 0.075 \\
 $\mu _{\frac{1}{2},\frac{1}{2}}(1405\to 1800) $& -0.143 & 0.035 & -0.006 & -0.114
   & -0.084 & 0.007 & -0.078 & -0.227 & -0.191 \\
 $\mu ^{\frac{1}{2},\frac{1}{2}}(1670\to 1670)$ & -0.802 & 0.116 & -0.030 & -0.715
   & 0.149 & -0.008 & 0.141 & -0.653 & -0.575 \\
 $\mu _{\frac{1}{2},\frac{1}{2}}(1670\to 1800)$ & 0.105 & 0.068 & 0.001 & 0.174 &
   0.201 & -0.011 & 0.190 & 0.305 & 0.364 \\
 $\mu _{\frac{1}{2},\frac{1}{2}}(1800\to 1800)$ & 0.676 & -0.223 & 0.030 & 0.484 &
   -0.053 & 0.003 & -0.049 & 0.623 & 0.434 \\
$ \mu_{\frac{3}{2},\frac{1}{2}}(1520\to 1520) $& -0.154 & -0.002 & -0.005 & -0.160
   & -0.058 & 0.007 & -0.051 & -0.212 & -0.211 \\
$ \mu_{\frac{3}{2},\frac{1}{2}}(1520\to 1690) $& -0.181 & 0.033 & -0.007 & -0.155
   & -0.087 & 0.006 & -0.080 & -0.268 & -0.236 \\
 $\mu_{\frac{3}{2},\frac{1}{2}}(1690\to 1690) $& 0.278 & -0.081 & 0.012 & 0.209 &
   0.132 & -0.007 & 0.125 & 0.410 & 0.334 \\
$ \mu_{\frac{3}{2},\frac{3}{2}}(1520\to 1520) $& -0.461 & -0.005 & -0.015 & -0.481
   & -0.173 & 0.021 & -0.153 & -0.635 & -0.634 \\
 $\mu_{\frac{3}{2},\frac{3}{2}}(1520\to 1690)$ & -0.543 & 0.098 & -0.021 & -0.466
   & -0.260 & 0.019 & -0.241 & -0.803 & -0.707 \\
 $\mu_{\frac{3}{2},\frac{3}{2}}(1690\to 1690)$ & 0.834 & -0.243 & 0.036 & 0.627 &
   0.396 & -0.021 & 0.375 & 1.230 & 1.000 \\
 \end{tabular}
 \end{ruledtabular}
\end{table}

In Table~\ref{magmomres} we show the results found for the magnetic moments with the two above mentioned models, where we have separated the contribution to the magnetic moment originating from the constituent quarks or valence quarks, sea quarks, etc. The magnetic moments obtained within the SU(6) quark model correspond to the summation of the contributions from the spin and orbital parts for the valence quarks, while in the chiral quark model case we also have contributions from the sea quarks, as discussed in the previous section.

As can be seen from the table, the major part of the contribution to the magnetic moment of the low-lying $\Lambda$ resonances comes from the valence quarks, but the consideration
of the sea quarks present in the $\chi$QM makes that the magnetic moment for all the elastic transitions decreases in magnitude (absolute value), except for the $\Lambda(1405)$ $1/2^-$,
whose magnetic moment augments. The magnetic moments of the $\Lambda(1405)$ and $\Lambda(1800)$ $1/2^-$ resonances are positive and have opposite sign
to the one associated with the $\Lambda(1670)$, whereas the magnetic moment of the $\Lambda(1520)$ $3/2^-$ is negative and with opposite sign to the corresponding one
for the $\Lambda(1690)$ $3/2^-$.

We can compare these results with the ones obtained within unitary chiral theories. In particular, in Ref.~\cite{JidoMM} the magnetic moments for the $\Lambda(1405)$ and $\Lambda(1670)$ resonances were calculated, getting the following values:

\begin{align}
\mu_{\frac{1}{2},\frac{1}{2}}(1405\to1405)&=+(0.2-0.5)\mu_N,\nonumber\\
\mu_{\frac{1}{2},\frac{1}{2}}(1670\to1670)&\sim -0.29\mu_N,\label{Uchipt}\\
\left|\mu_{\frac{1}{2},\frac{1}{2}}(1405\to1670)\right|&\sim 0.023\mu_N\nonumber.
\end{align}
The magnetic moment of the $\Lambda(1670)$ is negative, while the one for the $\Lambda(1405)$ is positive and the magnetic moment for the transition is much smaller in absolute value than the one of the $\Lambda(1670)$. These features are also shared by the SU(6) quark model and the chiral quark model. However, although in the chiral quark model, somehow, the effect of the meson cloud is taken into account when determining the magnetic moment of the resonances, the results obtained here are quite different as compared to the ones found within unitary chiral theories, which also considers the effect of the meson cloud. Comparing the results in Eq.~(\ref{Uchipt}) with the corresponding ones in Table~\ref{magmomres} for the chiral quark model, we see that the magnetic moment of the $\Lambda(1405)$ in the unitary chiral theories is at least two times bigger than the one obtained with the chiral quark model, while the magnetic moment of the $\Lambda(1670)$ is around a factor of 2 smaller than the one determined within the chiral quark model. For the transition magnetic moment between the $\Lambda(1405)$ and $\Lambda(1670)$,  the chiral quark model predicts a magnitude around 3 times bigger than the result related to unitary chiral theories. It is also interesting to notice that in the chiral quark model, the magnetic moment of the $\Lambda(1405)$ and the transition of this state to the $\Lambda(1670)$ are comparable in magnitude, while in the unitary chiral theories, the magnetic moment of the $\Lambda(1405)$ is, at least, about 10 times bigger than the one for its transition to the $\Lambda(1670)$.

Where do these differences in the magnetic moments come from? In the unitary chiral theories, the $\Lambda(1405)$ and $\Lambda(1670)$ get generated though the hadron-hadron dynamics involved in the interaction of the different coupled channels considered, like, $\bar K N$, $\pi \Sigma$, $\eta \Lambda$, etc., and its unitarization through the determination of the scattering matrix using the Bethe-Salpeter equation~\cite{JidoMM,Kaiser,OllerM,JidoL}. The result is that these states can be interpreted as a kind of molecule where the hadrons which form them keep their identities. From the other side, the chiral quark model considers the effect that the presence of the pseudoscalar mesons, $\pi$, $K$, $\eta$ and $\eta^\prime$, can originate in the magnetic moment through a quark fluctuation, as shown in Eqs.~(\ref{tran}) and ~(\ref{chiQL}). The different results obtained within the two models shows clearly the different nature that these states posses in the two formalism.

Definitively, the measurement of these magnetic moments or transition magnetic moments will help in clarifying the nature of these kind of resonances.

\section{Conclusions}
We have determined the magnetic moment for the low-lying $1/2^-$ and $3/2^-$ $\Lambda$ resonances, as well as their transitions, within a nonrelativistic SU(6) quark model and within the chiral quark model. In case of the chiral quark model, we have evaluated the contribution coming from the constituent or valence quarks as well as from the sea quarks obtained from the fluctuation process based on the emission of Goldstone bosons from the quarks. We have found that the major part of the magnitude associated to the magnetic moment of the $\Lambda$ states studied has its origin in the valence quarks, however, the contribution from the sea quarks give rise to an augment or reduction of them. For the case of the $\Lambda(1405)$ and $\Lambda(1670)$, we have also compared the results obtained within unitary chiral theories, in which these $\Lambda$ states get generated through the hadron dynamics, with our findings using the chiral quark model. Although some features, like the sign of the magnetic moment of these two $\Lambda$ states and its transition, are common in both models, the difference in magnitude reveals the different nature involved for these states within the two models. Measurements of these magnetic moments will be of great help in understanding the structure and nature of these resonances.

\section{Acknowledgments}
A.M.T and K.P.K are indebted to Prof. A. Hosaka for very useful discussions.
H.~Dahiya and N.~Sharma would like to thank DAE-BRNS (Ref No: 2010/37P/48/BRNS/1445) for the financial support.
A. M.T and K.P.K sincerely acknowledge the financial support from FAPESP and CNPq.


\begin{thebibliography}{99}
\bibitem{Cheng}
  T.~P.~Cheng and L.~-F.~Li,
  %``Why naive quark model can yield a good account of the baryon magnetic moments,''
  Phys.\ Rev.\ Lett.\  {\bf 80}, 2789 (1998).

  \bibitem{Linde}
  J.~Linde, T.~Ohlsson and H.~Snellman,
  %``Octet baryon magnetic moments in the chiral quark model with configuration mixing,''
  Phys.\ Rev.\ D {\bf 57}, 452 (1998).
%  [hep-ph/9709353].

%  [hep-ph/9709295].

\bibitem{JidoMM}
  D.~Jido, A.~Hosaka, J.~C.~Nacher, E.~Oset and A.~Ramos,
  %``Magnetic moments of the Lambda(1405) and Lambda(1670) resonances,''
  Phys.\ Rev.\ C {\bf 66}, 025203 (2002).
%  [hep-ph/0203248].

\bibitem{Chiang}
  W.~-T.~Chiang, S.~N.~Yang, M.~Vanderhaeghen and D.~Drechsel,
  %``Magnetic dipole moment of the S(11)(1535) from the gamma p ---\rangle gamma eta p reaction,''
  Nucl.\ Phys.\ A {\bf 723}, 205 (2003).
  %[nucl-th/0211061].

  \bibitem{Harleen}
  H.~Dahiya and M.~Gupta,
Phys. Rev. D {\bf 66}, 051501(R) (2002); Phys. Rev. {\bf D 67}, 114015 (2003); Phys. Rev. {\bf D
67}, 074001 (2003); Int. Jol. Mod.
Phys. A {\bf 19}, 5027 (2004); H. Dahiya, M. Gupta, and J.M.S. Rana,
Int. Jol. Mod. Phys. A {\bf 21}, 4255 (2006); H. Dahiya and M. Gupta, Phys. Rev. {\bf D 78}, 014001
(2008).

  \bibitem{Liu}
  J.~Liu, J.~He and Y.~B.~Dong,
  %``Magnetic moments of negative-parity low-lying nucleon resonances in quark models,''
  Phys.\ Rev.\ D {\bf 71}, 094004 (2005).

  \bibitem{Yu}
  L.~Yu, X.~-L.~Chen, W.~-Z.~Deng and S.~-L.~Zhu,
  %``Radiative decays of decuplet baryons, Lambda(1405) and Lambda (1520) hyperons,''
  Phys.\ Rev.\ D {\bf 73}, 114001 (2006).
%  [hep-ph/0602171].

  \bibitem{MiSo}
  M.~Doring, E.~Oset and S.~Sarkar,
  %``Radiative decay of the Lambda(1520),''
  Phys.\ Rev.\ C {\bf 74}, 065204 (2006).
 % [nucl-th/0601027].


  \bibitem{JidoFF}
  D.~Jido, M.~Doering and E.~Oset,
  %``Transition form factors of the N*(1535) as a dynamically generated resonance,''
  Phys.\ Rev.\ C {\bf 77}, 065207 (2008).
  %[arXiv:0712.0038 [nucl-th]].
  %%CITATION = ARXIV:0712.0038;%%

  \bibitem{Hideko}
  H.~Nagahiro, L.~Roca, A.~Hosaka and E.~Oset,
  %``Hidden gauge formalism for the radiative decays of axial-vector mesons,''
  Phys.\ Rev.\ D {\bf 79}, 014015 (2009).

  \bibitem{Daniel}
  D.~Gamermann, E.~Oset and B.~S.~Zou,
  %``The Radiative decay of psi(3770) into the predicted scalar state X(3700),''
  Eur.\ Phys.\ J.\ A {\bf 41}, 85 (2009).
%  [arXiv:0805.0499 [hep-ph]].

  \bibitem{Neetika}
  N.~Sharma, H.~Dahiya, P.~K.~Chatley and M.~Gupta,
  %``Spin 1/2^+, spin 3/2^+ and transition magnetic moments of low lying and charmed baryons,''
  Phys.\ Rev.\ D {\bf 81}, 073001 (2010).
%  [arXiv:1003.4338 [hep-ph]].

\bibitem{Geng1}
  L.~S.~Geng, J.~Martin Camalich, L.~Alvarez-Ruso and M.~J.~Vicente Vacas,
  %``Leading SU(3)-breaking corrections to the baryon magnetic moments in Chiral Perturbation Theory,''
  Phys.\ Rev.\ Lett.\  {\bf 101}, 222002 (2008).

  \bibitem{Geng2}
  L.~S.~Geng, J.~Martin Camalich and M.~J.~Vicente Vacas,
  %``Leading-order decuplet contributions to the baryon magnetic moments in Chiral Perturbation Theory,''
  Phys.\ Lett.\ B {\bf 676}, 63 (2009);
  %``Electromagnetic structure of the lowest-lying decuplet resonances in covariant chiral perturbation theory,''
  Phys.\ Rev.\ D {\bf 80}, 034027 (2009).

\bibitem{IK}
  N.~Isgur and G.~Karl,
  %``P Wave Baryons in the Quark Model,''
  Phys.\ Rev.\ D {\bf 18}, 4187 (1978),
    %``Positive Parity Excited Baryons in a Quark Model with Hyperfine Interactions,''
  Phys.\ Rev.\ D {\bf 19}, 2653 (1979)
  [Erratum-ibid.\ D {\bf 23}, 817 (1981)].
  %%CITATION = PHRVA,D18,4187;%%

   \bibitem{Kaiser}
  N.~Kaiser, T.~Waas and W.~Weise,
  %``SU(3) chiral dynamics with coupled channels: Eta and kaon photoproduction,''
  Nucl.\ Phys.\ A {\bf 612}, 297 (1997).
 % [hep-ph/9607459].

  \bibitem{OllerM}
  J.~A.~Oller and U.~G.~Meissner,
  %``Chiral dynamics in the presence of bound states: Kaon nucleon interactions revisited,''
  Phys.\ Lett.\ B {\bf 500}, 263 (2001).

%  [hep-ph/0011146].
  %%CITATION = HEP-PH/0011146;%%


\bibitem{JidoL}
  D.~Jido, J.~A.~Oller, E.~Oset, A.~Ramos and U.~G.~Meissner,
  %``Chiral dynamics of the two Lambda(1405) states,''
  Nucl.\ Phys.\ A {\bf 725}, 181 (2003).
 % [nucl-th/0303062].
  %%CITATION = NUCL-TH/0303062;%%

  \bibitem{MKO1}
  A.~Martinez Torres, K.~P.~Khemchandani, L.~S.~Geng, M.~Napsuciale and E.~Oset,
  %``The X(2175) as a resonant state of the phi K anti-K system,''
  Phys.\ Rev.\ D {\bf 78}, 074031 (2008).
%  [arXiv:0801.3635 [nucl-th]].
  %%CITATION = ARXIV:0801.3635;%%
\bibitem{MKO2}
  K.~P.~Khemchandani, A.~Martinez Torres and E.~Oset,
  %``The N*(1710) as a resonance in the pi pi N system,''
  Eur.\ Phys.\ J.\ A {\bf 37}, 233 (2008).
%  [arXiv:0804.4670 [nucl-th]].


\bibitem{Luis}
  L.~Alvarez-Ruso, J.~A.~Oller and J.~M.~Alarcon,
  %``On the phi(1020) f0(980) S-wave scattering and the Y(2175) resonance,''
  Phys.\ Rev.\ D {\bf 80}, 054011 (2009).
%  [arXiv:0906.0222 [hep-ph]].

\bibitem{Hyodo}
  T.~Hyodo, S.~I.~Nam, D.~Jido and A.~Hosaka,
  %``Magnetic moments of the N(1535) resonance in the chiral unitary model,''
  nucl-th/0305023.
  %%CITATION = NUCL-TH/0305023;%%

  \bibitem{Weinberg}
 S.~Weinberg,
  %``Phenomenological Lagrangians,''
  Physica A {\bf 96}, 327 (1979).
  %%CITATION = PHYSA,A96,327;%%

  \bibitem{Manohar}
  A.~Manohar and H.~Georgi,
  %``Chiral Quarks and the Nonrelativistic Quark Model,''
  Nucl.\ Phys.\ B {\bf 234}, 189 (1984).
  %%CITATION = NUPHA,B234,189;%%

  \bibitem{Capstick}
  S.~Capstick and W.~Roberts,
  %``Quark models of baryon masses and decays,''
  Prog.\ Part.\ Nucl.\ Phys.\  {\bf 45}, S241 (2000).
%  [nucl-th/0008028].



\bibitem{Sharma}
  N.~Sharma, H.~Dahiya and P.~K.~Chatley,
  %``Extraction of the CKM matrix element V(us) from the hyperon semileptonic decays,''
  Eur.\ Phys.\ J.\ A {\bf 44}, 125 (2010).
  %%CITATION = EPHJA,A44,125;%%

  \bibitem{Song}
  X.~Song, J.~S.~McCarthy and H.~J.~Weber,
  %``Nucleon spin flavor structure in SU(3) breaking chiral quark model,''
  Phys.\ Rev.\ D {\bf 55}, 2624 (1997).
  %[hep-ph/9702363].
  %%CITATION = HEP-PH/9702363;%%
\end{thebibliography}
\end{document}